\documentclass[superscriptaddress,aps,preprintnumbers,amsmath,showpacs,amssymb,nofootinbib,reprint]{revtex4-1}
\usepackage{booktabs} 
\usepackage{array} 
\usepackage[table,xcdraw]{xcolor} 
\usepackage{amsmath,amssymb}
\usepackage{graphicx}
\usepackage{fancyhdr}
\usepackage{bm, color}
\usepackage{slashed,amsthm,amsfonts,empheq}
\usepackage[caption=false]{subfig}
\usepackage{hyperref}
\usepackage{multirow}
\usepackage[left=2cm,right=2cm,top=2cm,bottom=2cm,a4paper]{geometry}
\usepackage{comment}

\begin{document}

\renewcommand{\figurename}{Fig.}
\renewcommand{\tablename}{Table.}
\newcommand{\Slash}[1]{{\ooalign{\hfil#1\hfil\crcr\raise.167ex\hbox{/}}}}
\newcommand{\bra}[1]{ \langle {#1} | }
\newcommand{\ket}[1]{ | {#1} \rangle }
\newcommand{\beq}{\begin{equation}}  \newcommand{\eeq}{\end{equation}}
\newcommand{\bef}{\begin{figure}}  \newcommand{\eef}{\end{figure}}
\newcommand{\bec}{\begin{center}}  \newcommand{\eec}{\end{center}}
\newcommand{\non}{\nonumber}  \newcommand{\eqn}[1]{\begin{equation} {#1}\end{equation}}
\newcommand{\laq}[1]{\label{eq:#1}}  
\newcommand{\dd}[1]{{d \o d{#1}}}
\newcommand{\Eq}[1]{Eq.~(\ref{eq:#1})}
\newcommand{\Eqs}[1]{Eqs.~(\ref{eq:#1})}
\newcommand{\eq}[1]{(\ref{eq:#1})}
\newcommand{\Sec}[1]{Sec.\ref{chap:#1}}
\newcommand{\ab}[1]{\left|{#1}\right|}
\newcommand{\vev}[1]{ \left\langle {#1} \right\rangle }
\newcommand{\bs}[1]{ {\boldsymbol {#1}} }
\newcommand{\lac}[1]{\label{chap:#1}}
\newcommand{\SU}[1]{{\rm SU{#1} } }
\newcommand{\SO}[1]{{\rm SO{#1}} }
\def\({\left(}
\def\){\right)}
\def\dt{{d \o dt}}
\def\diag{\mathop{\rm diag}\nolimits}
\def\Spin{\mathop{\rm Spin}}
\def\O{\mathcal{O}}
\def\U{\mathop{\rm U}}
\def\Sp{\mathop{\rm Sp}}
\def\SL{\mathop{\rm SL}}
\def\tr{\mathop{\rm tr}}
\def\ebq{\end{equation} \begin{equation}}
\newcommand{\OR}{~{\rm or}~}
\newcommand{\AND}{~{\rm and}~}
\newcommand{\EV}{ {\rm \, eV} }
\newcommand{\KEV}{ {\rm \, keV} }
\newcommand{\MEV}{ {\rm \, MeV} }
\newcommand{\GEV}{ {\rm \, GeV} }
\newcommand{\TEV}{ {\rm \, TeV} }
\def\o{\over}
\def\a{\alpha}
\def\b{\beta}
\def\c{\varepsilon}
\def\d{\delta}
\def\e{\epsilon}
\def\f{\phi}
\def\g{\gamma}
\def\h{\theta}
\def\k{\kappa}
\def\l{\lambda}
\def\m{\mu}
\def\n{\nu}
\def\p{\psi}
\def\q{\partial}
\def\r{\rho}
\def\s{\sigma}
\def\t{\tau}
\def\u{\upsilon}
\def\w{\omega}
\def\x{\xi}
\def\y{\eta}
\def\z{\zeta}
\def\D{\Delta}
\def\G{\Gamma}
\def\H{\Theta}
\def\L{\Lambda}
\def\F{\Phi}
\def\P{\Psi}
\def\S{\Sigma}
\def\me{\mathrm e}
\def\ol{\overline}
\def\tl{\tilde}
\def\*{\dagger}


\newcounter{HTQ}
\newcommand{\commentHT}[1]{{\bf\textcolor{red}{\stepcounter{HTQ}
[$\bullet\,$ \textcolor{black}{\theHTQ}\hspace*{0.15cm} #1]}}}


\begin{flushright}
{YITP-26-23}
\end{flushright}

\title{
Beyond thresholds: 
reconstructing 
UV 
physics from 
IR 
expansions
}

\author{
Hiromasa Takaura
}
\affiliation{
Center for Gravitational Physics and Quantum Information, Yukawa Institute for Theoretical Physics, Kyoto University,
Kyoto 606-8502, Japan
} 
\author{
Wen Yin
}
\affiliation{Department of Physics, Tokyo Metropolitan University, 
Minami-Osawa, Hachioji-shi, Tokyo 192-0397 Japan}

\begin{abstract}

We show that ultraviolet information can be extracted from low-energy expansion coefficients, 
assuming analyticity and the absence of massless singularities. 
By reorganizing the low-energy expansion through an inverse Laplace transform 
and a controlled coarse-graining procedure, 
we make ultraviolet behavior accessible beyond the cutoff of the effective field theory. 
In particular, we determine the sign of the beta function and the associated dynamical scale 
directly from the low-energy expansion of a physical observable below the mass thresholds
in QED and QCD-like theories.
\end{abstract}

\maketitle
\flushbottom

\vspace{1cm}

\section{Introduction}

It is now widely believed that the Standard Model (SM) is an effective field theory (EFT) of some ultraviolet (UV) completion, and a central goal of particle physics is to uncover this underlying UV theory,
a quest that has been pursued for several decades~\cite{Brivio:2017vri}.
Along with experimental efforts, 
theoretical developments in exploring the UV completion are underway~\cite{Georgi:1974sy,Nilles:1983ge,Polchinski:1994mb,Vafa:2005ui, Adams:2006sv,Albertus:2026fbe}.
Such theoretical studies can provide new perspectives that may offer clues to 
identifying new physics, and seem increasingly important, 
given that no clear experimental signals for new physics have been found so far.

It is often stated that low-energy physics is insensitive to its UV completion. 
This widely shared wisdom can be made precise in two different senses.
(a) In the strict low-energy limit, heavy degrees of freedom decouple from physical observables~\cite{Appelquist:1974tg}.
(b) Even if power corrections of $\mathcal{O}(Q^2/\Lambda^2)$ are included to arbitrarily high orders, one cannot predict a behavior at energies beyond the cutoff scale $\Lambda$~\cite{Wilson:1973jj,Weinberg:1978kz}.

In this work we challenge viewpoint (b). 
We establish a nontrivial connection between IR and UV physics, 
thereby showing that UV information can be extracted from IR expansion coefficients of a physical observable. 
By reorganizing the low-energy expansion through an inverse Laplace transform 
and a controlled coarse-graining procedure, implemented, e.g., 
via a numerical fit or a physically motivated analytic ansatz, 
we demonstrate that aspects of the high-energy behavior beyond the thresholds 
can be reconstructed.

{The inverse Laplace transform (often referred to as the Borel transform) is widely used in the analysis of factorially divergent perturbative series to extract nonperturbative information encoded in large-order behavior, such as instanton or renormalon effects (see, e.g.,\cite{Beneke:1998ui, Marino:2012zq}). 
Here we instead apply this transform to the low-energy expansion of an observable which has a finite radius of convergence, and use it to extract ultraviolet renormalization-group information beyond that convergence radius.}

{This work builds on Ref.~\cite{Takaura:2024bcj},  
where the inverse Laplace transform was applied to correlation functions 
in asymptotically free, mass-gapped theories. 
There it was shown that the transformed correlators admit an analytic structure 
that enables IR expansion coefficients to be extracted 
from UV information such as the OPE, 
including a controlled treatment of duality-violation effects. }
In contrast, the method proposed here relies on the analyticity of physical observables and provides 
a quantitative bottom-up approach, as explained below.

\section{General Framework for Accessing UV from IR}\label{sec:setup}

We consider a physical quantity $S$ that depends on a 
momentum transfer $q^2$ or $Q^2=-q^2$.
We assume, as generally expected, that $S(Q^2)$ is analytic everywhere except for singularities on the negative real axis \cite{Eden:1966dnq, Zwicky:2016lka}.
($q^2>0$ or $Q^2<0$ corresponds to the time-like region.)
We further assume that $S(Q^2)$ is analytic at $Q^2=0$ (no massless singularities).
Under these assumptions, $S(Q^2)$ admits the Taylor expansion around the origin,
\begin{equation}
    S(Q^2)=\sum_{n=0}^{\infty} c_n \left( \frac{Q^2}{Q_{\rm thr}^2}\right)^n.  \label{loweneexp}
\end{equation}
The closest singularity is located at $Q^2=-Q^2_{\rm thr}$, reflecting the existence of a massive particle
or a production threshold.
$Q_{\rm thr}^2$  effectively acts as a UV cutoff, as the above low-energy expansion converges only within $|Q^2| \leq Q_{\rm thr}^2$ at most.
$c_n$ with larger $n$ corresponds to higher-dimensional operators of the EFT.

While the series~\eqref{loweneexp} cannot access $|Q^2|>Q_{\rm thr}^2$,
we explore UV physics beyond $Q^2_{\rm thr}$
using low-energy expansion coefficients $\{c_0, c_1, c_2, \cdots \}$ as input.
We consider the inverse Laplace transform \cite{Takaura:2024bcj},
\begin{equation}
\tilde{S}(\tau)=\frac{1}{2 \pi i} \int_{-i \infty}^{i \infty} \frac{d z}{z} S(Q^2=1/z) e^{\tau z}
=\sum_{n=0}^{\infty} \frac{c_n}{n!} \left( \frac{\tau}{Q_{\rm thr}^2}\right)^n .
    \label{invLaplace}
\end{equation}
The integration contour in the complex  $z(=1/Q^2)$-plane 
is depicted in Fig.~\ref{fig:contours}.
We note that the inverse Laplace transform is well defined 
even without employing the series expansion of $S(Q^2)$.
The variable $\tau$ is newly introduced and has mass dimension two.
It can be proven that the radius of convergence of the series expansion in Eq.~\eqref{invLaplace} is infinite
due to the factorial suppression $1/n!$ in Eq.~\eqref{invLaplace}, in sharp contrast to Eq.~\eqref{loweneexp}.
Therefore, by increasing $n_{\rm max}$, the truncated approximation
$\tilde{S}(\tau)|_{n_{\rm max}}=\sum_{n=0}^{n_{\rm max}} (c_n/n!) (\tau/Q^2_{\rm thr})^n$ 
becomes accurate over an increasingly wide energy range of $\tau$.

{Nevertheless, the exact inverse relation,
\begin{equation}
S(Q^2)=\frac{1}{Q^2}\int_{0}^{\infty} d\tau\, \tilde{S}(\tau)\, e^{-\tau/Q^2},
\label{exactinverse}
\end{equation}
shows that substituting $\tilde{S}(\tau)|_{n_{\rm max}}$ 
simply reproduces the original low-energy expansion and therefore does not
yield a prediction for $Q^2 \gg Q_{\rm thr}^2$. 
This reflects the one-to-one correspondence between the Laplace and inverse Laplace transforms.
To go beyond the original cutoff, a ``deformation'' of 
$\tilde{S}(\tau)|_{n_{\rm max}}$ is required.
By this we mean a {\it coarse-graining} procedure,
namely, constructing an analytic or numerical function,
that approximates the IR series for $\tilde{S}$ within its validity range
and can be smoothly and consistently extrapolated to the UV region,
so that Eq.~\eqref{exactinverse} yields nontrivial UV information. 
}
The range of $\tau$ where $\tilde{S}(\tau)|_{n_{\rm max}}$ is reliable 
is determined by comparing successive truncations,
$\tilde{S}(\tau)|_{n_{\rm max}}$ and $\tilde{S}(\tau)|_{n_{\rm max}-1}$.
This range typically extends up to 
$\tau/Q_{\rm thr}^2 = \mathcal{O}(n_{\rm max})$.
Within this validity range, we fit 
$\tilde{S}(\tau)|_{n_{\rm max}}$ 
or project it onto a physically motivated UV ansatz.\footnote{The ansatz is constrained, for instance, by the operator product expansion~\cite{Wilson:1972ee}, which suggests a power-like behavior at large $Q^2$, $S(Q^2)\sim (Q^2/Q_{\rm thr}^2)^p \times ({\text{Wilson coeff.}})$
(or equivalently $\tilde{S}(\tau)\sim (\tau/Q_{\rm thr}^2)^p \times (\text{Wilson coeff.})$ at large $\tau$).}
Owing to the exponential suppression in the kernel $e^{-\tau/Q^2}$,
the UV extrapolation contributes only mildly to $S(Q^2)$,
unless the coarse-grained approximation reproduces the IR series
to parametrically high accuracy (see Fig.~\ref{fig2}).

In summary, the main idea of this work consists 
of the following three steps:
\begin{description}
\item[i] exploit the good convergence of the inverse Laplace--transformed quantity $\tilde{S}(\tau)$\footnote{For our approach to work, other inversible transformation may work as well.}
 \item[ii]  construct a coarse-grained $\tilde{S}(\tau)$ valid with arbitrary large $\tau$ from the information in the range $\tau \lesssim n_{\rm max} Q_{\rm thr}^2$
\item[iii] extract UV information such as $S(Q^2)$ at $Q^2 > Q_{\rm thr}^2$. 
\end{description}

We now demonstrate its power using two representative examples.

\begin{figure}
\begin{center}
\includegraphics[width=7.5cm]{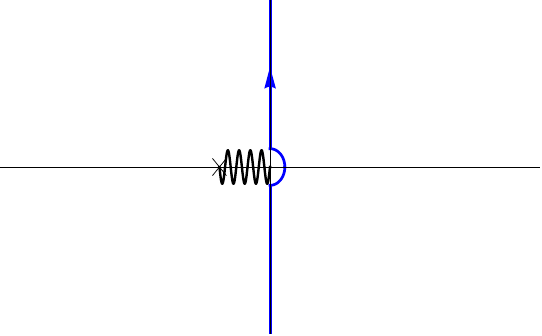}
\end{center}
\caption{Integration contours in the complex $z$-plane.}
\label{fig:contours}
\end{figure}

\section{From IR to UV in QED}
\label{sec:QED}

As a first example, we consider one-flavor QED.
We assume access only to the low-energy expansion below the electron pair-production threshold.
For $Q \lesssim m_e$, the electron is integrated out, 
and the remaining degrees of freedom are represented solely by the massless photon
(the Euler-Heisenberg theory),
where the coupling does not run 
due to the decoupling of the electron.

Let us consider
\begin{equation}
S(Q^2) \equiv \alpha(m_e^2) (1+\Pi(Q^2;\mu^2=m_e^2)) , \laq{Sq}
\end{equation}
where $\Pi$ is the one-loop photon vacuum polarization.
Its low-energy expansion reads
\begin{align}
& S(|Q^2|  < 4 m_e^2)
=\sum_{n=0}^{\infty} c_n \left(\frac{Q^2}{Q_{\rm thr}^2} \right)^n \nonumber \\ 
&=\alpha(m_e^2)+\frac{4 \alpha^2(m_e^2)}{15 \pi} \frac{Q^2}{4 m_e^2}-\frac{4 \alpha^2(m_e^2)}{35 \pi} \left( \frac{Q^2}{4 m_e^2} \right)^2+\cdots , \label{SloweneQED}
\end{align}
where $Q^2_{\rm thr}=4 m_e^2$ and $\alpha(m_e^2)=1/137$ 
(and thus $c_n$ does not run).
As expected, this expansion fails to describe the behavior beyond the threshold;
see the dotdashed and dashed lines in Fig.~\ref{fig2}.

We consider the inverse Laplace transform~\eqref{invLaplace}, where $\tilde S$ is constructed from the coefficients $c_n$.
The validity range of the truncated approximation to $\tilde{S}(\tau)$ expands as more terms are included,
 in contrast to the low-energy expansion of $S(Q^2)$ (see Appendix~\ref{sec:SysErrorAn}).

{
To extract UV information, we employ two approaches for step~(ii). 
The first approach is to fit $\tilde S$ and then perform the Laplace transformation back to $S$ to get the UV behavior. 
}

{To this end, and without loss of generality, we employ the \texttt{Interpolation} function in \textit{Mathematica} to fit $\tilde S$.\footnote{
The function performs piecewise-polynomial interpolation between successive data points. 
The option \texttt{InterpolationOrder} specifies the degree of the local interpolating polynomials: 
$n=0$ corresponds to a piecewise-constant (step) function, 
$n=1$ to piecewise-linear interpolation, 
and higher values ($n=2,\dots,5$ in the present analysis) produce progressively smoother interpolants with increasing continuity of derivatives.}
The result is shown in Fig.~\ref{fig2}, for interpolation orders $1,3,5,7,9,$ and $10$. 
Here we take $n_{\rm max}=10$ and fit the output of $\tilde S$ in the range $\tau \in [0,\tau_{\rm max}]$ with $\tau_{\rm max}=10 m_e^2$, using a linear spacing of $m_e^2/5$. 
Since the Laplace integral \eqref{exactinverse} extends beyond the fitting range, 
the fitted function is extrapolated.\footnote{
When the resulting \texttt{InterpolatingFunction} is evaluated outside its nominal domain, 
the same local polynomial used in the outermost interval is analytically continued beyond the endpoint. 
Thus, the extrapolation employed here corresponds to a direct continuation of the endpoint polynomial.}}

{Small interpolation orders yield results close to the exact solution 
(black dot-dashed line) and {can clearly} deviate from the IR expansion 
(black dotted line).
In contrast, as the interpolation order approaches $n_{\rm max}(=10)$, 
the result tends to reproduce the original truncated IR polynomial 
(see the red curve overlapping with the dotted curve). 
An excessively accurate fit that merely reconstructs the original high-degree polynomial 
therefore fails to implement genuine ``coarse-graining" and leads to the original unstable behavior of $S$.
}

More generally, the rapid growth of $\tilde S$ at large $\tau$, 
induced by high-degree polynomials, 
translates into poor convergence in the reconstructed $S$ after the Laplace transform.
A viable coarse-graining procedure should suppress this large-$\tau$ growth relative to the original truncated polynomial. 
Once this condition is satisfied, 
the detailed form of the extrapolation becomes only mildly relevant, 
provided that the fit accurately reproduces the data within $\tau \lesssim n_{\rm max} m_e^2$.

For much larger $n_{\rm max}$, the accessible UV range of $S$ becomes wider, 
and even a case of multiple thresholds can be handled (see Appendix~\ref{app:discrimination}). \\

As an alternative, we introduce a well-motivated assumption about the UV theory.
{This approach allows us to work with smaller $n_{\rm max}$
and to directly extract UV properties such as the sign of the beta function and the associated dynamical scale.}
We assume that the UV theory belongs to a class of weakly coupled theories 
governed by a single coupling, denoted by $\alpha_{\rm UV}$.
In this case, standard OPE arguments suggest that, at high energies,
$S(Q^2) \sim \alpha_{\rm UV}^k(Q^2)$ with some $k>0$,
where the leading behavior scales as $\mathcal{O}(Q^0)$,
consistent with the dimensionality of $S(Q^2)$.
This single-coupling assumption can be tested a posteriori, as discussed below.
Neither $\alpha_{\rm UV}$ nor $k$ is specified a priori.

\begin{figure}
\includegraphics[width=8.5cm]{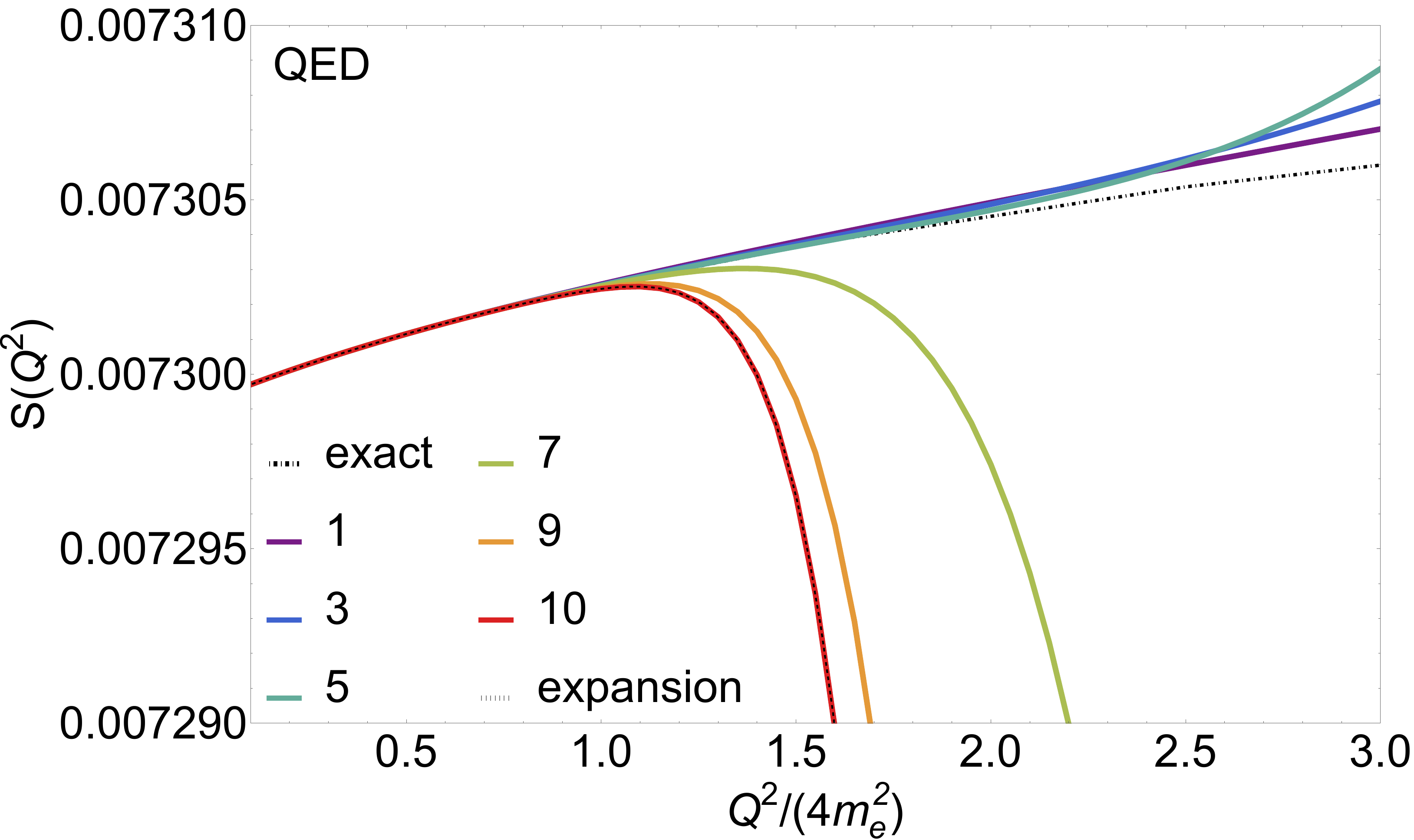}
\caption{
The UV behavior of $S$ reconstructed from the IR expansion 
by fitting $\tilde S$ (solid lines) in the QED model. 
We take $\tau_{\rm max}=10 m_e^2$ and $n_{\rm max}=10$.  
The colors denote the interpolation orders ($1,3,5,7,9,10$). 
The exact result is shown as a black dot-dashed line, 
while the IR expansion is shown as a dotted-dashed line, 
which overlaps with the red solid curve for interpolation order $10$.
}
\label{fig2}
\end{figure}

Under the above assumption, it is possible to
study the beta function of the UV theory. 
The beta function turns out to be related to $\tilde{S}(\tau)$ by
\begin{equation}
    \frac{d \ln \tilde{S}}{d \ln \tau}=k \beta_{\tilde{S}}(\alpha_{\tilde{S}})/\alpha_{\tilde{S}}  \quad{} \text{for $\tau \gg Q^2_{\rm thr}$}, \label{dStl}
\end{equation}
{where $\alpha_{\tilde{S}} \equiv [(\tau/Q_{\rm thr}^2)^{-p} \tilde{S}(\tau)  /a_0]^{1/k}$ is the effective coupling with $p$ being the classical scaling of $\tl S$ (see Appendix\,\ref{sec:weaksinglegeneral}).} 
This is because the large-$\tau$ behavior corresponding to $S(Q^2) \sim  \a_{\rm UV}^k(Q^2)$
is given by $\tilde{S}(\tau) \sim  \a_{\rm UV}^k(\tau)$
by noting that \cite{Hayashi:2023fgl}
\begin{equation}
   \frac{1}{2 \pi i} \int_{-i \infty}^{i \infty} \frac{dz}{z}  z^a e^{\tau z}
   =\frac{1}{\pi} \left( \frac{1}{\tau} \right)^a \Gamma(a) \sin(\pi a) ,
    \label{invLapformula}
\end{equation}
and that $\tilde{S}(\tau)$ is renormalization-group (RG) invariant following from the RG invariance of $S(Q^2)$.
Actually, Eq.~\eqref{dStl} holds beyond the leading order in perturbation theory 
and the subscript $\tilde{S}$ represents a scheme of defining $\alpha_{\rm UV}$.
A more detailed discussion, including the case of different mass dimension of $S$, is given in Appendix.~\ref{sec:weaksinglegeneral}.

Examining the left-hand side of Eq.~\eqref{dStl} using the given $c_n$
reveals the sign of the beta function, i.e., whether the UV theory is asymptotically free or not, since $k$ must be positive as required by perturbativity. 
Figure~\ref{fig:QED_betafn} shows $d\ln \tilde S|_{n_{\rm max}}/d\ln \tau$
as a function of $\tau/(4m_e^2)$,
identifying the validity range.
The beta function is found to be positive for $\tau \gg 4 m_e^2$, 
correctly reproducing that QED is not asymptotically free.
We note that the sign of the beta function is scheme independent 
in the weak-coupling regime, and the fact that the magnitude 
of eq.~\eqref{dStl}, which is of order $\mathcal{O}(\alpha_{\tilde{S}})$,
is much smaller than unity supports that we are indeed in such a regime.

Furthermore, using the one-loop running 
$\alpha_{\tilde S}(\tau)=1/[b_0\ln(\tau/\Lambda_{\tilde S}^2)]$, 
Eq.~\eqref{dStl} leads to an estimate of the dynamical scale,
\begin{equation}
    \ln(\Lambda_{\tilde{S}}^2/m_e^2) \simeq 
    \frac{k}{\frac{d \ln \tilde{S}}{d \ln \tau}} 
    + \ln(\tau/m_e^2) .
    \label{dynamicalscale}
\end{equation}
Here $\Lambda_{\tilde{S}}$ denotes the dynamical scale {(or Landau pole scale)} of the UV theory and 
$b_0$ is the one-loop beta-function coefficient. 
Notably, the coefficient $b_0$
 cancels out in this expression.
From the result for $\tilde{S}(\tau)|_{n_{\rm max}=8}$ at $\tau/(4 m_e^2)=3$,
we obtain $\ln(\Lambda_{\tilde{S}}^2/m_e^2) \approx 1.5 \times 10^3\, k \gtrsim 1.5 \times 10^3$.
In general, the dynamical scale $\Lambda_{\tilde{S}}$ defined in different schemes may differ by a factor of $\mathcal{O}(1)$; namely, 
$\Lambda_{\tilde{S}} = c\, \Lambda_{\overline{\rm MS}}$ with $c=\mathcal{O}(1)$. 
The above estimate is rough, with a relative uncertainty of order $10\%$, 
but it nevertheless provides a sensible order-of-magnitude result. 
For comparison, the one-loop result in the $\overline{\rm MS}$ scheme is 
$\ln(\Lambda_{\overline{\rm MS}}^2/m_e^2)=1291$.

\begin{figure}
\begin{center}
\includegraphics[width=8cm]{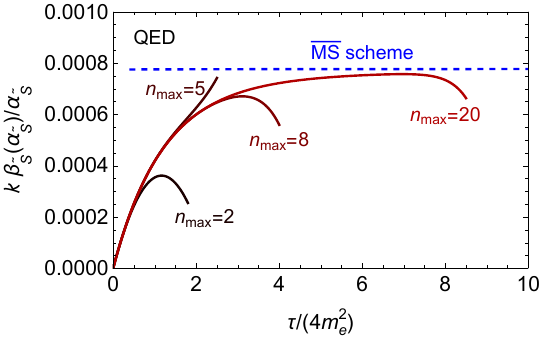}
\caption{$d \ln \tilde{S}|_{n_{\rm max}}/d \ln \tau$. See Eq.~\eqref{dStl}.
The truncation order of the low-energy expansion is indicated by $n_{\rm max}$.
The curves are cut off considering their validity ranges.
The $\overline{\rm MS}$-scheme result for $k \beta(\alpha)/\alpha = k/\log(\Lambda_{\rm QED}^2/\tau)$ with $k=1$ is also shown as the blue dashed line.}
\label{fig:QED_betafn}
\end{center}
\end{figure}

Next we extract $S(Q^2)$ at high energies, $Q^2 \gg 4 m_e^2$.
As a reference point, we take $n_{\rm max}=6$
and fit $\tilde{S}(\tau)|_{n_{\rm max}=6}$ 
in the large-$\tau$ region within its validity range
with
$\tilde{S}(\tau)|_{\rm fit} = \text{const.} + \text{const.} \times \log \tau$,
consistent with the above UV ansatz,
as clarified in Appendix~\ref{sec:weaksinglegeneral}.
The reconstructed $S(Q^2)$ is shown in Fig.~\ref{fig:QED_est_SQsq}
(see Appendix~\ref{sec:SysErrorAn} for technical details).
It reproduces the exact result of the UV theory
well beyond the threshold of the original low-energy expansion valid at
$Q^2/(4 m_e^2)<1$.
This allows for a good fit with smaller $n_{\rm max}$
than in the case without the UV assumption shown in Fig.~\ref{fig2}.
The single-coupling nature of the UV theory can be further examined
by checking whether the corresponding quantities derived from
different observables exhibit consistent behavior.

\begin{figure}
\begin{center}
\includegraphics[width=8.5cm]{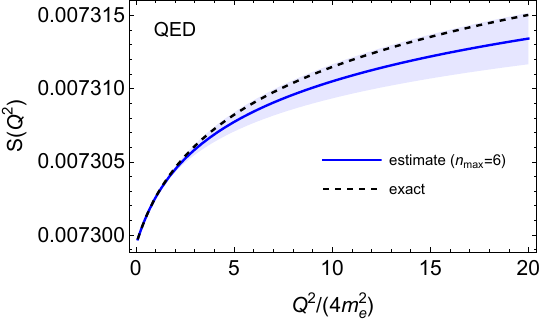}
\caption{Extracted $S(Q^2)$ (blue) with an error band, compared with the exact result (dashed) obtained in the UV theory.}
\label{fig:QED_est_SQsq}
\end{center}
\end{figure}

\section{QCD-like theory}
\label{sec:CPN}

We now demonstrate that our method also works for a theory whose IR dynamics are governed nonperturbatively, in contrast to the previous example. 
We consider the $\mathbb{C}P^{N-1}$ model in two-dimensional spacetime in the large-$N$ limit~\cite{DAdda:1978vbw}, 
which shares key features with QCD, such as asymptotic freedom and a confining potential that gives rise to mesonic bound states. {We again assume a single-weak coupling in the UV for simplicity, while the numerical fit as coarse-graining procedure without the assumption works as well. }

We consider the Fourier transform of the static potential, $V_{\rm mom}(Q^2)$,
\begin{align}
V(R)=-\frac{4 \pi}{N} \int_{-\infty}^{\infty} \frac{d Q}{2\pi} \frac{1}{Q^2} V_{\rm mom}(Q^2) e^{i Q R} .
\end{align}
The mass dimension of $V_{\rm mom}(Q^2)$ is two. We therefore define a dimensionless quantity,
$S(Q^2) \equiv V_{\rm mom}(Q^2)/m^2$,
normalized by the dynamical scale $m$.
The threshold is given by $Q_{\rm thr}^2=4 m^2$.
Our analysis proceeds in a manner largely parallel to the previous example. 
In this case, taking into account the dimensionality of $V_{\rm mom}(Q^2)$, the high-energy behavior is expected to be
$S(Q^2) \sim (Q^2/m^2)\,\alpha_{\rm UV}^k(Q^2)$,
which leads to some differences in the subsequent analysis.
Further details are provided in Appendix.~\ref{sec:SysErrorAn}.

From the low-energy coefficients $c_n$, 
we study the beta function through 
\begin{equation}
     \frac{d \ln \tilde{S}}{d \ln \tau}-1=k \beta_{\tilde{S}}(\alpha_{\tilde{S}})/\alpha_{\tilde{S}} . \label{betafnCPN}
\end{equation}
Fig.~\ref{fig:CPN_betafn} correctly indicates that the beta function is negative (at $\tau \gg 4m^2$).
This implies that elementary particles different from ``observed particles'' (mesons) 
could exist in the UV theory, and also that perturbative prescription becomes
valid at high energies.
The successful extraction of $S(Q^2)$ is demonstrated in Appendix.~\ref{sec:SysErrorAn}.
These results indicate that the present method provides a bridge between the low-energy nonperturbative regime and the weakly coupled UV regime. 

\begin{figure}
\begin{center}
\includegraphics[width=8cm]{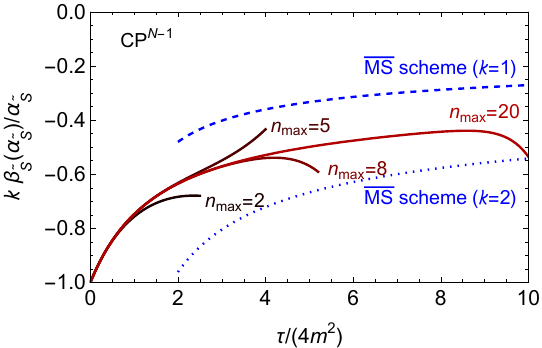}
\caption{$d \ln \tilde{S}|_{n_{\rm max}}/d \ln \tau-1$. See Eq.~\eqref{betafnCPN}.
The truncation order of the low-energy expansion is indicated by $n_{\rm max}$.
The curves are cut considering their validity ranges.
The $\overline{\rm MS}$ scheme result is also shown by the blue dashed line for $k=1$ and $k=2$. }
\label{fig:CPN_betafn}
\end{center}
\end{figure}

\section{Outlook}

Let us provide possible applications of the proposal of this work. 

For instance, 
our approach to extracting UV physics from given IR information can be combined with the
argument~\cite{Pham:1985cr, Ananthanarayan:1994hf, Adams:2006sv}, 
which constrains IR effective field theories from expected UV features. 
The inverse Laplace transform has an alternative integral representation for $\tau>0$,
\begin{equation}
    \tilde{S}(\tau)=\frac{1}{\pi} \int_0^{-1/Q^2_{\rm thr}} \frac{dz}{z} {\rm Im} \, S(Q^2=1/(z+i0)) e^{\tau z} ,
\end{equation}
obtained via a contour deformation, when $S(Q^2)$ converges fast enough in the limit $Q^2 \to \infty$.
Although the positivity argument can determine the sign of each low-energy expansion coefficient~\cite{Pham:1985cr, Ananthanarayan:1994hf, Adams:2006sv},
the present argument globally constrains the sign of $\tilde{S}(\tau)$ at any $\tau>0$,
when the sign of ${\rm Im} \, S(Q^2=1/(r+i0))$ is (nonperturbatively) available based on
the dispersion relation.
Furthermore, since our method can connect $c_n$ with UV dynamics through Eq.~\eqref{invLaplace},
we could obtain strong constrains on $c_n$.
We will discuss such applications in our future publication.

\acknowledgments
 This work was supported by JSPS KAKENHI Grant Nos. JP19K14711(H.T.),
JP23K13110(H.T.), 20H05851 (W.Y.), 21K20364 (W.Y.), 22K14029 (W.Y.), and 22H01215 (W.Y.), and Incentive Research Fund for Young Researchers from Tokyo Metropolitan University (W.Y.).  
H.T. is the Yukawa Research Fellow supported by Yukawa Memorial
Foundation.
\clearpage
\appendix 

\onecolumngrid 

\section{Supplementary results on $S(Q^2)$}

\subsection{QED example}

$\Pi(Q^2)$ at one loop is given by
\begin{equation}
    \Pi(Q^2)=\frac{2 \alpha(\mu^2)}{\pi} \int_0^1 dx \, x(1-x) \ln \left[\frac{x(1-x)Q^2+m_e^2}{\mu^2} \right] ,
\end{equation}
from which the exact result of $S(Q^2)$ in Figs.~\ref{fig:QED_est_SQsq} and \ref{fig:higherorderQED} is given. 
The first few terms of the low-energy expansion of $S(Q^2)$ is given by Eq.~\eqref{SloweneQED}, whereas the high-energy expansion $Q^2 \gg m_e^2$ yields
\begin{align}
S(Q^2)
&=\alpha(\mu^2)\left (1+\frac{\alpha(\mu^2)}{3 \pi} \log(Q^2 e^{-5/3}/\mu^2) \right) \\ \nonumber
&\quad{}+\mathcal{O}(m_e^2/Q^2) . \label{SQsqasymptotic}
\end{align}
Therefore the parameter $k$ in the main text and Eq.~\eqref{ptStilde} is given by $k=1$.
The beta-function of QED at one loop is given by
\begin{equation}
\mu^2 \frac{d}{d \mu^2} \alpha(\mu^2)=\frac{\alpha^2(\mu^2)}{3 \pi} ,
\end{equation}
and then the running coupling is given by
\begin{equation}
    \alpha(\mu^2)=\frac{3 \pi}{\log(\Lambda_{\rm QED}^2/\mu^2)} .
\end{equation}
$\Lambda_{\rm QED}$ is determined in the units of the electron mass by
$\alpha(\mu^2=m_e^2)=1/137$.

\subsection{$\mathbb{C}$P$^{N-1}$ example}

The action of this theory is given by
\begin{equation}
    S=\frac{N}{\lambda_0} \int d^2 x  [D_{\mu} z^\dagger D_{\mu} z+f (z^{\dagger} z-1)] ,
\end{equation}
where $z$ is a complex $N$ component vector, $D_{\mu}=\partial_{\mu}+i A_{\mu}$, and 
$\lambda_0$ is the (bare) 't Hooft coupling.
$A_{\mu}$ and $f$ are auxiliary fields, whose kinetic terms are absent in the original classical action.
This theory is solvable in the $1/N$ expansion
and known to exhibit confinement;
a linear potential arises from 
the dynamical photon, whose kinetic term emerges due to quantum corrections.
As a result, low-energy degrees of freedom are ``mesons,'' made of fundamental scalars. 
The dynamical scale of this asymptotically free theory is denoted by $m$ 
and the scalars acquire dynamical mass of $m$.

The exact result of $V_{\rm mom}(Q^2)$ in the large-$N$ limit is known to be~\cite{DAdda:1978vbw}
\begin{align}
& V_{\rm mom}(Q^2) \nonumber \\
&=\left[\frac{2 \sqrt{Q^2+4 m^2}}{Q^2 \sqrt{Q^2}} \log \left(\frac{\sqrt{Q^2+4 m^2}+\sqrt{Q^2}}{\sqrt{Q^2+4 m^2}-\sqrt{Q^2}} \right)-\frac{4}{Q^2}  \right]^{-1} .
\end{align}
The low-energy expansion of $S(Q^2)$ is
\begin{align}
&S(Q^2)=\sum_{n=0}^{\infty} c_n \left(\frac{Q^2}{4 m^2} \right)^n  \nonumber \\
&=3+\frac{6}{5} \frac{Q^2}{4 m^2}-\frac{36}{175} \left(\frac{Q^2}{4 m^2} \right)^2+\cdots , \label{CPNpotlowene}
\end{align}
which converges for $|Q^2| < 4 m^2$, where the threshold is given by 
twice the scalar mass, $(2m)^2$.
The high-energy expansion is given by
\begin{align}
    S(Q^2)
    &=\frac{Q^2}{m^2} \bigg[\frac{\lambda}{8 \pi}+\frac{\lambda^2}{16 \pi^2} \left(1+\frac{1}{2} \ln(\mu^2/Q^2) \right) +\mathcal{O}(\lambda^3)  \nonumber \\
    &\qquad{}\quad{}+\mathcal{O}(m^2/Q^2) \bigg]
\end{align}
Therefore, the parameters in Eq.~\eqref{ptStilde}
correspond to $p=1$ and $k=1$.
The beta function is given by
\begin{equation}
    \mu^2 \frac{d}{d \mu^2} \lambda=-\frac{\lambda^2}{4 \pi} ,
\end{equation}
whose solution is given by
\begin{equation}
    \lambda(\mu^2)=\frac{4 \pi}{\log(\mu^2/m^2)} .
\end{equation}

\section{General discussion for single weak coupling theories}

\label{sec:weaksinglegeneral}

Suppose that an observable $S(Q^2)$ is given in perturbation theory 
in a single weak coupling at high energies $Q^2/Q_{\rm thr}^2 \gg 1 $. 
\begin{align}
S(Q^2)
  &=(Q^2/Q_{\rm thr}^2)^p \nonumber \\
  &\quad{} \times [a_0 \alpha_{\rm UV}^{k}(Q^2)+a_1 \alpha_{\rm UV}^{k+1}(Q^2)+\mathcal{O}(\alpha_{\rm UV}^{k+2}) \nonumber \\
  &\qquad{}+\mathcal{O}(Q^2_{\rm thr}/Q^2) ] ,  \label{ptS}
\end{align}
where $p$ is an integer usually determined by the classical scaling of $S(Q^2)$ 
and $k$ is a positive integer.
Using the {relation~\eqref{invLaplace}
}, 
it follows that $\tilde{S}(\tau)$ is also be given in perturbation theory:
\begin{align} 
  \tilde{S}(\tau)
  &=(\tau/Q_{\rm thr}^2)^p \nonumber \\
  &\quad{}\times [\tilde{a}_0 \alpha_{\rm UV}^{k}(\tau)+\tilde{a}_1 \alpha_{\rm UV}^{k+1}(\tau)+\mathcal{O}(\alpha_{\rm UV}^{k+2}(\tau)) \nonumber \\
  &\qquad{}+\mathcal{O}(Q^2_{\rm thr}/\tau) ] ,
  \label{ptStilde}
\end{align}
where $\tilde{a}_0 , \tilde{a}_1$ can be expressed in terms of $a_0, a_1$ \cite{Hayashi:2023fgl}.

From $\tilde{S}$, one can define a coupling and the corresponding beta function:
\begin{equation}
    \alpha_{\tilde{S}} \equiv [(\tau/Q_{\rm thr}^2)^{-p} \tilde{S}(\tau)  /a_0]^{1/k} , \label{alphaSdef}
\end{equation}
\begin{align}
   & \tau \frac{d}{d \tau} \alpha_{\tilde{S}}(\tau)
    \equiv \beta_{\tilde{S}}(\alpha_{\tilde{S}}) \nonumber \\ 
  & =-b_0 \alpha^2_{\tilde{S}}(\tau)-b_1 \alpha^3_{\tilde{S}}(\tau)-b_{2,\tilde{S}} \alpha^4_{\tilde{S}}(\tau) \nonumber  \\ 
  &\quad{}+ \mathcal{O}(\alpha^5_{\tilde{S}}(\tau)) 
   +\mathcal{O}(Q^2_{\rm thr}/\tau) .  \label{betafn}
\end{align}
This is considered a renormalization scheme defined through $\tilde{S}$,
as it is related to another coupling $\alpha_{\rm UV}$ via
$\alpha_{\tilde{S}}(\mu^2)=\alpha_{\rm UV}(\mu^2)+\mathcal{O}(\alpha_{\rm UV}^2(\mu^2))$.
From these, we obtain
\begin{align}
   \frac{d \ln \tilde{S}}{d \ln \tau}-p=k \beta_{\tilde{S}}(\alpha_{\tilde{S}})/\alpha_{\tilde{S}} . \label{betafnfromtildeS}
\end{align}
It is known that the first two coefficients $b_0$ and $b_1$ do not depend on schemes~\cite{weinberg_1996}.
Therefore, the sign of the beta function is scheme independent,
as long as the coupling is weak $\alpha \lesssim 1$ at $\tau \gg Q_{\rm thr}^2$.

Eq.~\eqref{ptS} also determines the fit function to be used for large-$\tau$.
Rewriting the running coupling at $Q$ in terms of that at $\mu$, one obtains
\begin{align}
    &S(Q^2) \nonumber \\
    &=(Q^2/Q_{\rm thr}^2)^p   \nonumber \\
    &\quad{} \times [a_0 \alpha_{\rm UV}^{k}(\mu^2)+(a_1-a_0 b_0 k \log(Q^2/\mu^2)) \alpha_{\rm UV}^{k+1}(\mu^2) \nonumber \\
    &\qquad{}+\mathcal{O}(\alpha_{\rm UV}^{k+2})]  \label{pertseriesinmuSQsq} .
\end{align}
and correspondingly
\begin{align}
    &\tilde{S}(\tau) \nonumber \\
    &=(\tau/Q_{\rm thr}^2)^p \nonumber \\
    &\quad{} \times [a_0 \alpha_{\rm UV}^{k}(\mu^2)+(a_1-a_0 b_0 k \log(\tau e^{\gamma_E}/\mu^2)) \alpha_{\rm UV}^{k+1}(\mu^2)
   \nonumber\\
   &\qquad{}+\mathcal{O}(\alpha_{\rm UV}^{k+2}) ]  \label{pertseriesinmuStilde} ,
\end{align}
leading to the fit function of the form $const. + const. \times \log{\tau}$.
The parameters $x$ and $y$ introduced for the QED example below correspond to
\begin{equation}
    x \approx a_0 \alpha_{\rm UV}^{k}(\mu^2), \quad{} y \approx -a_0 b_0 k \alpha_{\rm UV}^{k+1}(\mu^2) .
\end{equation}

\section{Supplementary analyses}
\label{sec:SysErrorAn}

\subsection{QED example}

$\tilde{S}$ is shown as a function of $\tau$ in the bottom panel of Fig.~\ref{fig:2}. 
The region of convergence of $\tilde{S}(\tau)$ gets larger as more terms are added, in contrast to the low-energy expansion of $S(Q^2)$.
The validity range of expansions
is proportional to $n_{\rm max}$, as mentioned.

We give detailed explainations on how we obtain Fig.~\ref{fig:QED_est_SQsq} in the main part. 
We start with a fit of $\tilde{S}(\tau)|_{n_{\rm max}}$,
where the largest accessible $\tau$ region is focused
to safely neglect power corrections of $\mathcal{O}(4 m_e^2/\tau)$.
We identify the validity range of $\tilde{S}(\tau)|_{n_{\rm max}=6}$ by 
comparing it with the previous-order result $\tilde{S}(\tau)|_{n_{\rm max}=5}$,
and perform a fit in $2 \leq \tau/(4 m_e^2) \leq 3$ with 
$\tilde{S}(\tau)|_{\rm fit} = x + y \log(\tau e^{\gamma_E}/m_e^2)$ as default;
see Fig.~\ref{fig:QED_est_Stilde}.
We consider the following sources of systematic uncertainty.
(a) Fit range: we vary the lower end of the fit range to estimate the effect of neglected power corrections.
(b) Truncation order $n_{\rm max}$: the previous-order result ($n_{\rm max}-1$) is fitted using the same fit range,
thereby accounting for the uncertainty in $\tilde{S}(\tau)|_{n_{\rm max}}$, especially near the upper end of the range.
Among these, source (b) provides the dominant uncertainty.
The orange curve in Fig.~\ref{fig:QED_est_Stilde} shows the fit function corresponding to case (b).
We can predict $S(Q^2)$ using fit functions, as explained in the main text.
Use of two fit lines for the integrand $\tilde{S}(\tau)$ for large-$\tau$
gives an estimate of the uncertainty, 
which is shown by the blue band in Fig.~\ref{fig:QED_est_SQsq}.

\begin{figure}
\begin{center}
\includegraphics[width=8cm]{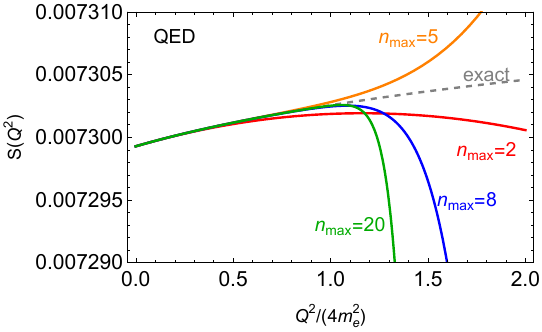}\\
\includegraphics[width=8cm]{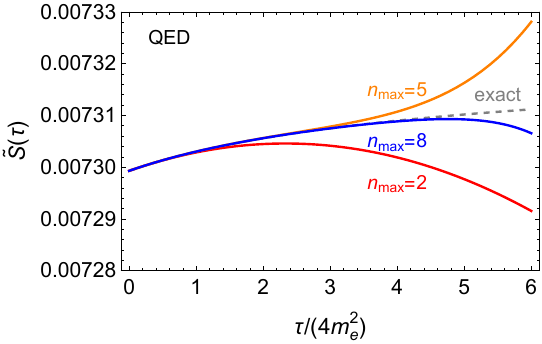}
\end{center}
\caption{The low-energy expansions of $S(Q^2)$ (top) and $\tilde{S}(\tau)$ (bottom), compared with the exact results.}
\label{fig:2}
\end{figure}

\begin{figure}
\begin{center}
\includegraphics[width=10cm]{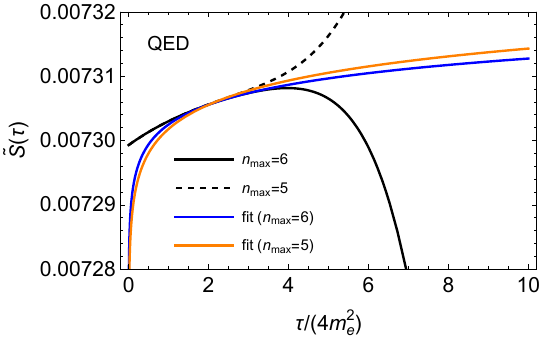}
\caption{Fits to $\tilde{S}(\tau)|_{n_{\rm max}}$ in a range 
$2 \leq \tau/(4 m_e^2) \leq 3$. See the text for details.}
\label{fig:QED_est_Stilde}
\end{center}
\end{figure}

We also give estimates on $S(Q^2)$ with different $n_{\rm max}$ in Fig.~\ref{fig:higherorderQED}, 
following the strategy explained in the main text and here. 
One can see that the estimates agree with the 
exact results with progressively reducing errors
as one goes to larger $n_{\rm max}$.
\begin{figure}
\includegraphics[width=7cm]{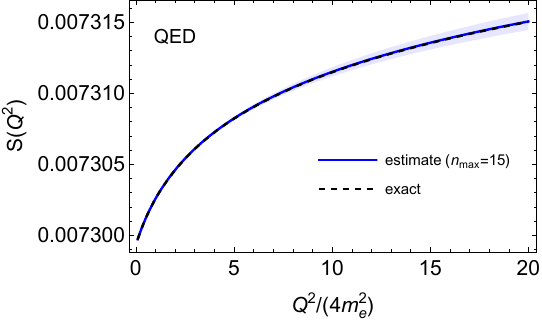}
\includegraphics[width=7cm]{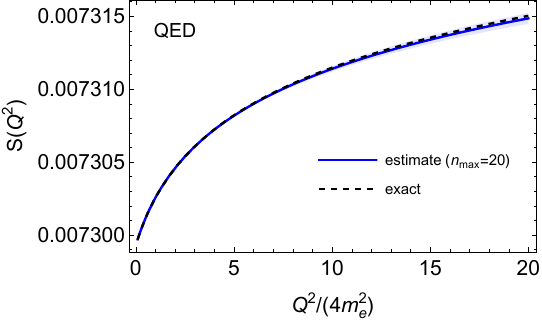}
\caption{Estimates of $S(Q^2)$ for the QED example with different truncation orders.
The band represents the systematic uncertainty.}
\label{fig:higherorderQED}
\end{figure}

\subsection{$\mathbb{C}$P$^{N-1}$ example}

This section provides supplementary explanations and results
for the $\mathbb{C}$P$^{N-1}$ example.
Fig.~\ref{fig:CPN_expansions} compares the low-energy expansions in 
the $Q^2$- and $\tau$-space, which again shows that the former 
is not valid beyond the threshold whereas the latter extends its validity range as larger $n_{\rm max}$.

\begin{figure}
\includegraphics[width=8cm]{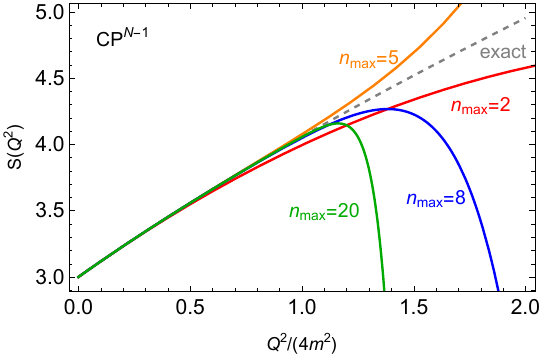}\\
\includegraphics[width=8cm]{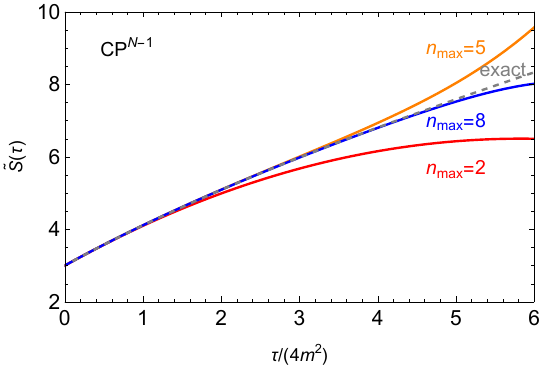}
\caption{The low-energy expansions of $S(Q^2)$ (top) and $\tilde{S}(\tau)$ (bottom), 
compared with the exact results, in the $\mathbb{C}$P$^{N-1}$ theory.}
\label{fig:CPN_expansions}
\end{figure}

Next, we explain how to extract $S(Q^2)$.
We perform a fit for Eq.~\eqref{betafnCPN} 
rather than $\tilde{S}(\tau)$ in this case.
In contrast to the QED case, it is suggested that 
the coupling is not very small in the accessible range; see Fig.~\ref{fig:CPN_betafn}.
Therefore, the same strategy, fitting $\tilde{S}(\tau)|_{n_{\rm max}}$ 
with the same formula (const.+log), is not valid sufficiently;
higher logarithmic terms can be important. However, introducing multiple fit parameters 
destabilized the fit significantly.
This is why we pursue a different strategy from the previous case.

In this case, 
$-k/[\log(\tau/\Lambda^2_{\tilde{S}})]$
serves a reasonable fit function for Eq.~\eqref{betafnCPN}, 
considering the one-loop running.
While $k$ is unknown a priori, we make the following observation
to constrain it.
We know that in asymptotically free theories with the mass gap,
the mass of (most of) composite particles is connected to the dynamical scale,
and thus, $\Lambda^2_{\tilde{S}}=\mathcal{O}(Q^2_{\rm thr})=\mathcal{O}(m^2)$.
Comparing the observed lines in Fig.~\ref{fig:CPN_betafn}
with $-k/[\log(\tau/(4 m^2))+c]$ with an $\mathcal{O}(1)$ constant $c$
thus can suggest a reasonable range of $k$.
The $\overline{\rm MS}$ sheme results (blue lines) 
can be regarded as one of those, as they are $-k//[\log(\tau/m^2)]$.
The figure thus suggests that $k=1$ or $k=2$.

We then fit Eq.~\eqref{betafnCPN}
with $-k/[\log(\tau/m^2)+c]$ with a fit parameter $c$,
where we choose either $k=1$ or $k=2$.
Note that we have only one fit parameter,
leading to stable fits.
As a benchmark point, we consider $n_{\rm max}=8$, where the line is valid up to $\tau/(4 m^2) \approx 4$.
In asymptotic free theories, going to higher energies than the threshold
can suppress power corrections and also make perturbation theory valid.
The fit range is taken to be $3 \leq \tau/(4 m^2) \leq 4$.
In Fig~\ref{fig:betafnfit}, we show the fit functions with $k=1$ (blue) or $2$ (orange) assumed.

\begin{figure}
    \centering
    \includegraphics[width=8cm]{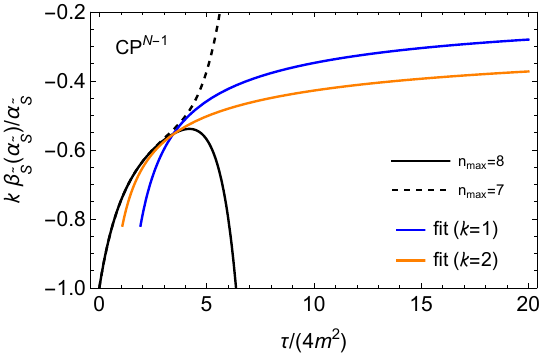}
    \caption{Fits to Eq.~\eqref{betafnCPN} with $n_{\rm max}=7$ or $8$
    in a range $3 \leq \tau/(4 m^2) \leq 4$. See the text for details.}
    \label{fig:betafnfit}
\end{figure}

\begin{figure}
\includegraphics[width=7cm]{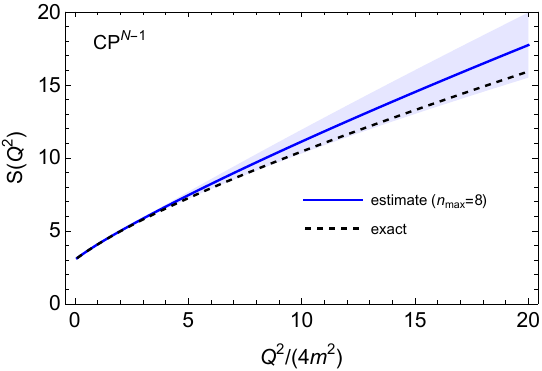}
\includegraphics[width=7cm]{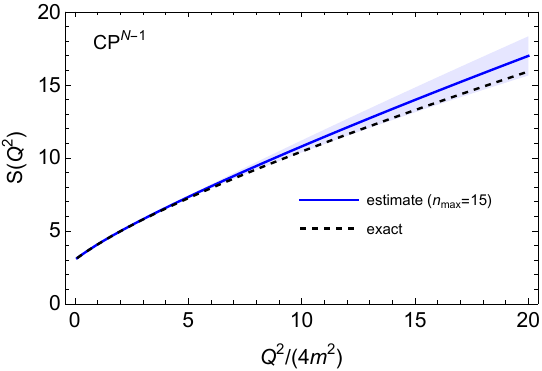}
\includegraphics[width=7cm]{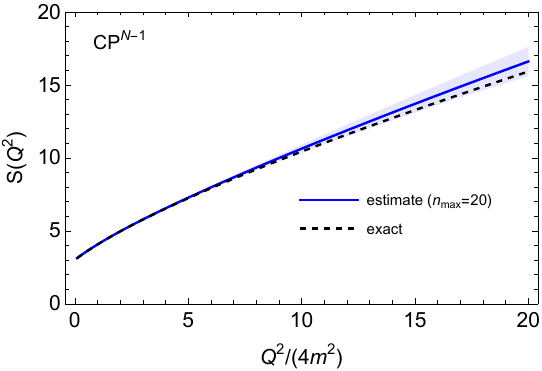}
\caption{Estimates of $S(Q^2)$ for the $\mathbb{C}$P$^{N-1}$ example with different truncation orders.}
\label{fig:higherorderCPN}
\end{figure}

Once a fit function is obtained with an explicit number $c$, 
by solving 
\begin{equation}
    \frac{d \ln \tilde{S}}{d \ln \tau}-1=-\frac{k}{\log(\tau/m^2)+c} ,
\end{equation}
we obtain
\begin{equation}
    \tilde{S}(\tau)=const. \times \tau \left( \ln{(\tau/m^2)}+c \right)^{-k} .
\end{equation}
The constant is adjusted at the upper end of the fit range.
For the integrand $\tilde{S}(\tau)$ in the inverse formula~\eqref{exactinverse},
we use the above function above the upper end of the fit range
whereas $\tilde{S}(\tau)|_{n_{\rm max}}$ below that point.

In estimating systematic uncertainties, 
the choice between $k=1$ and $2$ gives the dominant uncertainty, 
while 
we also consider the above mentioned systematic uncertainties (a) and (b) as in the QED case.
The results with $n_{\rm max}=8, 15, 20$ are shown in Fig.~\ref{fig:higherorderCPN}
with the the error bands determined by the dominant uncertainty.

\section{Beyond multi-thresholds: from IR to UV in QED}
\label{app:discrimination}
To illustrate the impact of our framework, 
we show that multiple thresholds can be resolved within our approach if $n_{\rm max}$ is large enough. 
To this end, we consider two-flavor QED with two leptons, 
``electron" $e$ and ``muon" $\mu$, of different masses $m_e$ and $m_\mu$ 
with $m_\mu \gtrsim m_e$. 

We define
\beq 
S(Q^2) \equiv \alpha(m_e^2) \left(1+\Pi(Q^2;\mu^2=m_e^2)\right), 
\laq{Sq2f}
\eeq 
with
\beq 
\Pi(Q^2)=\frac{2 \alpha(\mu^2)}{\pi}
\left(
\int_0^1 dx \, x(1-x) 
\ln \left[\frac{x(1-x)Q^2+m_e^2}{\mu^2} \right]
+
\int_0^1 dx \, x(1-x) 
\ln \left[\frac{x(1-x)Q^2+m_\mu^2}{\mu^2} \right]
\right).
\eeq 
We assume that $m_\mu$ is not extremely larger than $m_e$, 
so that higher-order running effects are negligible. 
In this observable, the lowest EFT threshold is again 
$Q_{\rm thr}^2=4m_e^2$, 
while higher-dimensional operators are modified by integrating out 
the muon at $Q^2\sim m_\mu^2$.

To demonstrate that our procedure reproduces the UV behavior even above the muon threshold, 
we perform steps (i)--(iii) described in the main text, 
without assuming a single weak coupling. 
That is, we first perform the inverse Laplace transform of $S$, 
fit $\tilde S(\tau)$, 
and then apply the Laplace transform. 
The result is shown by the colored solid lines in Fig.~\ref{fig2f}. 

We impose the renormalization condition 
$\lim_{Q^2\to 0}S^{\rm 2flavor}(Q^2)
=\lim_{Q^2\to 0}S^{\rm 1flavor}(Q^2)$. 
We take $n_{\rm max}=300$, which fully covers the muon threshold 
$\sim \O(10) m_e^2$, 
and choose $\tau_{\rm max}=300 m_e^2$ as the fitting range. 
The fitted function is extrapolated to larger $\tau$ 
to perform the Laplace transform.

For comparison, we also apply the same procedure directly to the conventional IR expansion 
(dashed lines with the same color). 
When the fit range is restricted to $Q^2 < 4 m_e^2$, 
where the series converges well, 
the reconstruction fails above the electron threshold and does not capture the higher-$Q^2$ behavior. 
This conclusion persists even if the fitting range is extended beyond the threshold, 
as the convergence of the IR expansion deteriorates.

We therefore conclude that the multi-thresholds can be resolved within our framework 
provided that $n_{\rm max}$ is sufficiently large. 
Such a resolution is not achievable by a na\"ive fit to {the IR expansion}
alone.
\begin{figure}
\includegraphics[width=12cm]{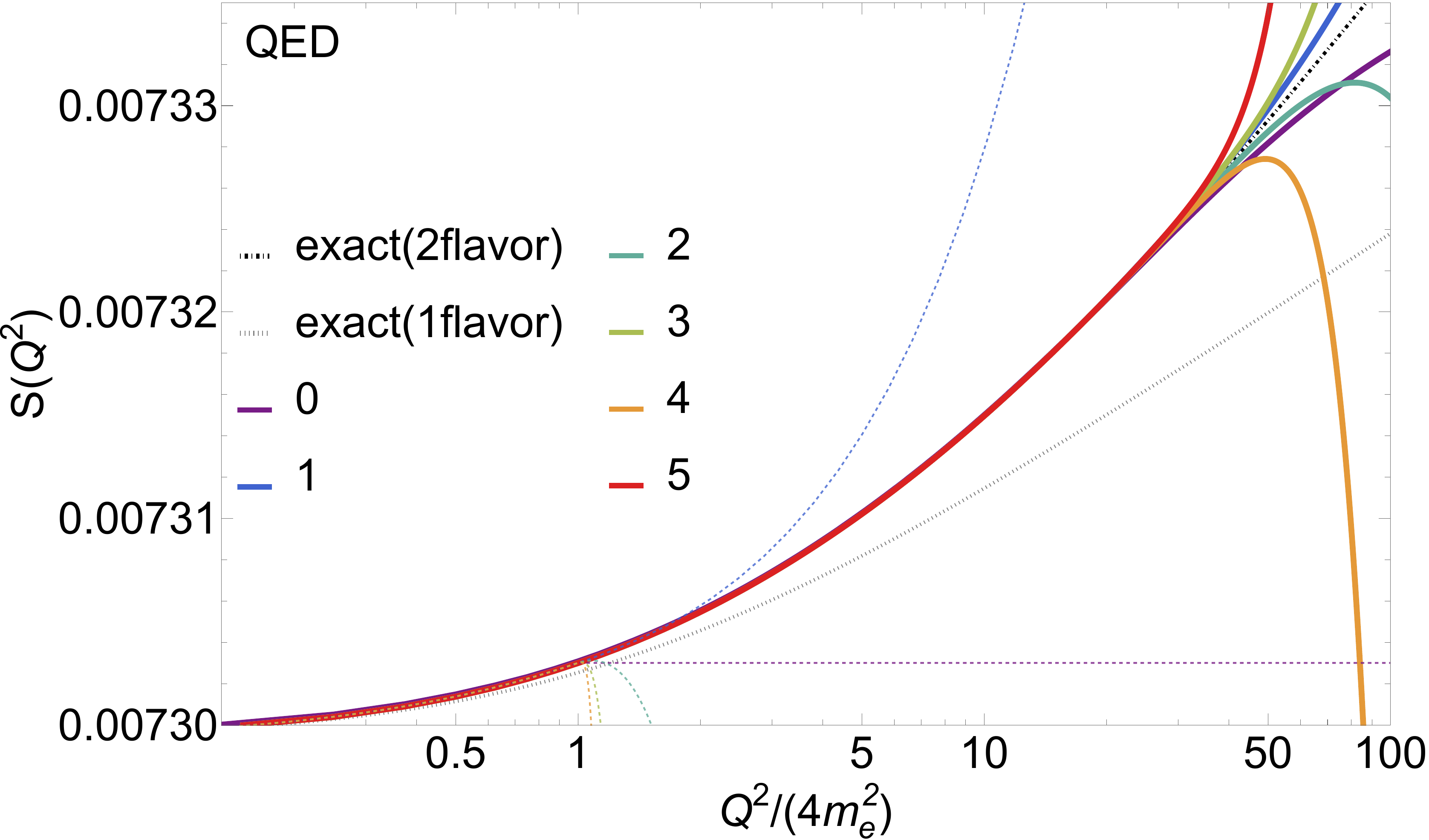}
\caption{
UV reconstruction of $S$ from the IR expansion 
via fitting $\tilde S$ (solid lines) in two-flavor QED. 
We take $m_\mu/m_e=3$, $\tau_{\rm max}=300 m_e^2$, and $n_{\rm max}=300$.  
Colors indicate interpolation orders ($0\cdots 5$). 
Dashed curves of the same color show the fit of the IR series 
($n_{\rm max}=300$) restricted to $Q^2<4m_e^2$. 
The black dot-dashed curve denotes the exact two-flavor result, 
and the black dotted curve the single-flavor case.
}
\label{fig2f}
\end{figure}

\bibliography{IRTOUV}

\end{document}